\documentclass[conference]{IEEEtran}
\usepackage{array, calc, color, multicol}
\usepackage{algorithm}
\usepackage[noend]{algorithmic} %, algorithm2e}
\usepackage{graphics, epstopdf, graphicx, epsfig, psfrag, epsf, subfigure}
\usepackage{amssymb, amsfonts, amsmath, times}
\usepackage{relsize}
\usepackage{multicol,balance, verbatim}
\usepackage{textcomp, mathrsfs, stmaryrd, setspace}
\usepackage{amssymb}
\newcommand{\ie}{{\em i.e., }}

\newcommand{\eg}{{\em e.g., }}
\newcommand{\Eg}{{\em E.g., }}

\newcommand{\ncrn}{{\tt{NCMI-Batch}}}
\newcommand{\ncin}{{\tt{NCMI-Instant}}}

\newtheorem{theorem}{Theorem}

\newtheorem{proposition}[theorem]{Proposition}

\newtheorem{definition}{Definition}
\newtheorem{example}{Example}

\DeclareGraphicsExtensions{.eps, .pdf, .png, .jpg}   % optional

\newcommand{\Nset}{\mathcal{N}}
\newcommand{\Mset}{\mathcal{M}}
\newcommand{\Hset}{\mathcal{H}}

\newcommand{\Wset}{\mathcal{W}}

\IEEEoverridecommandlockouts

\begin{document}
%\title{Prioritization in IDNC for Completion Time Reduction or Distortion Minimization}
%\title{On the Performance Bounds of Hybrid-interface Cooperative Network}
\title{Network Coding for Cooperative Mobile Devices with Multiple Interfaces}

%On the Performance of Network Coding for Cooperative Mobile Devices with Multiple Interfaces
%On the Minimum Number of Transmissions in Single-Hop Wireless Coding Networks
%On coding for cooperative data exchange

%\author{Yasaman Keshtkarjahromi, Hulya Seferoglu, Rashid Ansari\\
%{\small ECE Department, University of Illinois at Chicago}\\
%{ \small \tt ykesht2@uic.edu, hulya@uic.edu, ransari@uic.edu}\\
%\and
%Ashfaq Khokhar\\
%{\small ECE Department, Illinois Institute of Technology}\\
%{ \small \tt ashfaq@iit.edu }\\
%}

\author{Yasaman Keshtkarjahromi, Hulya Seferoglu, Rashid Ansari\\
{\small University of Illinois at Chicago}\\
{ \small \tt ykesht2@uic.edu, hulya@uic.edu, ransari@uic.edu}\\
\and
Ashfaq Khokhar\\
{\small Illinois Institute of Technology}\\
{ \small \tt ashfaq@iit.edu }\\
}

\maketitle

{$\hphantom{a}$}\vspace{-30pt}{}

\allowdisplaybreaks

%\vspace{-40pt}
\begin{abstract}
Cooperation among mobile devices and utilizing multiple interfaces such as cellular and local area links simultaneously are promising to meet the increasing throughput demand over cellular links. In particular, when mobile devices are in the close proximity of each other and are interested in the same content, device-to-device connections such as WiFi-Direct, in addition to cellular links, can be utilized to construct a cooperative system. However, it is crucial to understand the potential of network coding for cooperating mobile devices with multiple interfaces. In this paper, we consider this problem, and (i) develop network coding schemes for cooperative mobile devices with multiple interfaces, and (ii) characterize the performance of network coding by using the number of transmissions to recover all packets as a performance metric.

\end{abstract}

\section{\label{sec:introduction}Introduction}
The increasing popularity of diverse applications in today's mobile devices introduces higher demand for throughput, and puts a strain especially on cellular links. In fact, cellular traffic is growing exponentially and it is expected to remain so for the foreseeable future \cite{cisco_index}, \cite{ericsson_report}.

\begin{figure}[t!]
\vspace{-10pt}
\centering
\subfigure[\scriptsize The default operation for transmitting data from the core network to mobile devices]{ \scalebox{.3}{\includegraphics{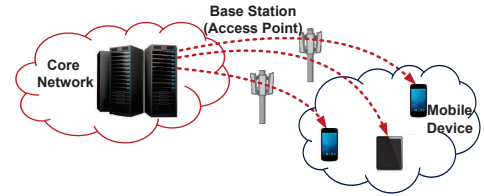}} } \hspace{5pt}
\subfigure[\scriptsize Cooperative mobile devices with multiple interfaces ]{ \scalebox{.3}{\includegraphics{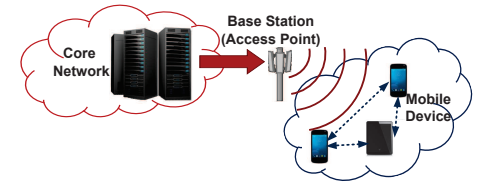}} }
\vspace{-10pt}
\caption{\scriptsize (a) The default operation in today's cellular systems: Each mobile device receives its data via unicast transmission over a cellular link. (b) Cooperative mobile devices with multiple interfaces: Mobile devices can cooperate and use multiple interfaces such as cellular and WiFi simultaneously to efficiently utilize available resources.
}
\vspace{-15pt}
\label{fig:intro_example}
\end{figure}

The default operation for transmitting data in today's networks is to connect each mobile device to the Internet via its cellular or WiFi connection, Fig.~\ref{fig:intro_example}(a). On the other hand, cooperation among mobile devices and utilizing multiple interfaces such as cellular and local area links simultaneously are promising to meet the increasing throughput demand. In particular, when mobile devices are in the close proximity of each other and are interested in the same content, device-to-device connections such as WiFi-Direct or Bluetooth can be opportunistically used to construct a cooperative system \cite{microcast}, \cite{microcast_allerton}, Fig.~\ref{fig:intro_example}(b). Indeed, this scenario is getting increasing interest \cite{microcast}. \Eg a group of friends may be interested in watching the same video on YouTube, or a number of students may participate in an online education class \cite{microcast}. However, it is crucial to understand the performance of cooperative mobile devices with multiple interfaces so that scarce wireless resources are efficiently utilized.

\begin{figure}[t!]
\begin{center}
\subfigure[\scriptsize Broadcasting four packets; $p_1$, $p_2$, $p_3$, $p_4$]{ \scalebox{.4}{\includegraphics{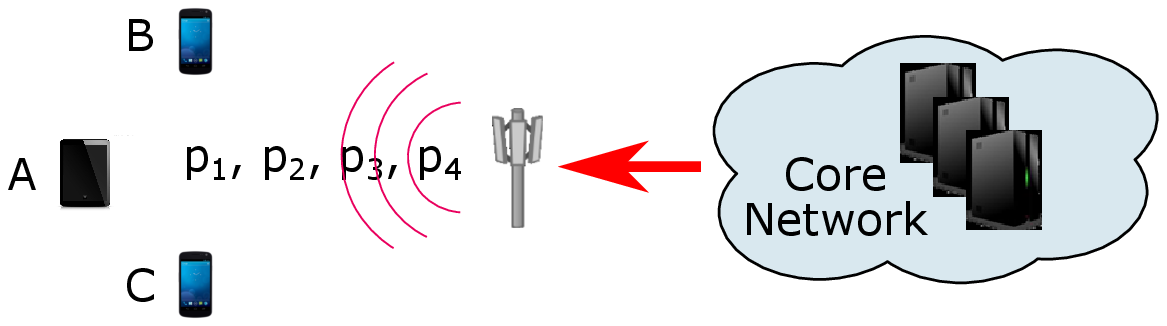}} } %
\subfigure[\scriptsize Missing packets after broadcast]{ \scalebox{.4}{\includegraphics{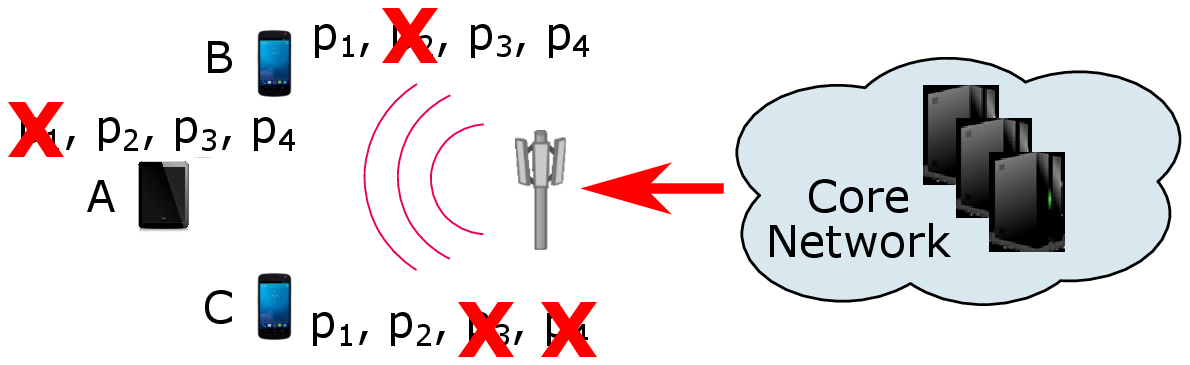}} }
\end{center}
\begin{center}
\vspace{-10pt}
\caption{\label{fig:missing} \scriptsize  Example scenario with packet losses. (a) Packets $p_1$, $p_2$, $p_3$, and $p_4$ are broadcast from the base station. (b) After the broadcast, $p_1$ is missing at mobile device $A$, $p_2$ is missing at $B$, and $p_3$ and $p_4$ are missing at $C$. }
\vspace{-18pt}
\end{center}
\end{figure}

In this paper, our goal is to develop a network coding \cite{nc_first}, \cite{cope} scheme for cooperative mobile devices with multiple interfaces operating simultaneously. In particular, we consider a scenario that a group of cooperative mobile devices, exploiting both cellular and local area links and within the proximity of each other, are interested in the same content, \eg video. In this setup, a common content is broadcast over cellular links\footnote{\scriptsize Note that broadcasting over cellular links is part of LTE \cite{3gpp_lte_broadcast}, \cite{ericsson_lte_broadcast}, \cite{qualcomm_lte_broadcast}, and getting increasing interest in practice, so we consider broadcast scenario instead of unicast.}, Fig. \ref{fig:missing}(a). However, mobile devices may receive only a partial content due to packet losses over cellular links, Fig. \ref{fig:missing}(b). The remaining missing content can then be recovered by utilizing both cellular and local area links simultaneously in a cooperative manner. In this setup, thanks to using different parts of the spectrum, cellular links and local area links operate concurrently. Thus, a mobile device can receive two packets simultaneously; one via cellular, and the other via local area links. The fundamental question in this context, and the focus of this paper, is to design and develop efficient network coding algorithms that take into account cooperation among mobile devices and multiple interfaces.

The performance of network coding in single-interface systems has been considered in previous work, \cite{cope}, \cite{cc_wang}, \cite {salim_broadcast}, \cite{RouayhebITW10}, \cite{RouayhebISIT10}, \cite{SprintsonQShine10}, \cite{TajbakhshAusCTW14}, \cite{parastoo_broadcast}, in the context of broadcasting a common content over cellular links, and repairing the missing content via (i) retransmissions over cellular links, or (ii) by exploiting local area device-to-device connections. The following example demonstrates the potential of network coding in single-interface systems.

\begin{example} \label{example_indv_xmit}
Let us consider Fig. \ref{fig:missing}(a), where four packets, ${p_1,p_2,p_3,p_4}$ are broadcast from the base station. Assume that after the broadcast, $p_1$ is missing at mobile device $A$, $p_2$ is missing at $B$, and $p_3$ and $p_4$ are missing at $C$, Fig. \ref{fig:missing}(b). The missing packets can be recovered via re-transmissions (broadcasts) over cellular links. Without network coding, four transmissions are required so that each mobile device receives all the packets. With network coding, two transmissions from the base station are sufficient: $p_1+p_2+p_3$ and $p_4$. After these two transmissions, all mobile devices have the complete set of packets. As can be seen, network coding reduces four transmissions to two, which shows the benefit of network coding in this setup.

Now let us consider packet recovering by exploiting local area links. Assume again that after the broadcast, $p_1$ is missing at mobile device $A$, $p_2$ is missing at $B$, and $p_3$ and $p_4$ are missing at $C$. Without network coding, four transmissions are required to recover all missing packets in all mobile devices. With network coding in the local area, two transmissions are sufficient: (i) mobile device $B$ broadcasts $p_1+p_3$, and (ii) $A$ broadcasts $p_2+p_4$. After these two transmissions, all mobile devices have all the packets. In this example, by taking advantage of network coding, the number of transmissions are reduced from four to two transmissions.
\hfill $\Box$
\end{example}

The above example demonstrates the benefit of network coding when a single interface is used. However, mobile devices can exploit multiple interfaces including cellular and local area links simultaneously. The following example demonstrates the potential of network coding in this setup.

\begin{example}
Let us consider Fig. \ref{fig:missing}(b) again, and assume that after the broadcast, $p_1$ is missing at device $A$, $p_2$ is missing at $B$, and $p_3$ and $p_4$ are missing at $C$. When both cellular and local area links are exploited, the following transmissions are simultaneously made to recover the missing packets: (i) the base station broadcasts $p_1+p_3$ via cellular links, and (ii) mobile device $A$ broadcasts $p_2+p_4$ via local area links. As can be seen, the number of transmission slots is reduced to one transmission slot from two as compared to Example~\ref{example_indv_xmit}.
\hfill $\Box$
\end{example}

Thus, mobile devices with multiple interfaces and cooperation have potential of improving throughput significantly. However, it is crucial to understand and quantify the potential of network coding for cooperating mobile devices with multiple interfaces. In this paper, we consider this problem, and (i) develop network coding schemes, namely {\em network coding for multiple interfaces (NCMI)}, for cooperative mobile devices with multiple interfaces, and (ii) characterize the performance of the proposed network coding schemes, where we use packet completion time, which is the number of transmission slots to recover all packets, as a performance metric. The following are the key contributions of this work:
\begin{itemize}
\item We develop a lower bound on the packet completion time when network coding is employed by cooperative mobile devices with multiple interfaces.
\item We propose a network coding algorithm; \ncrn, where packets are network coded as a batch to improve the throughput of cooperative mobile devices with multiple interfaces. By taking into account the number of packets that each mobile device would like to receive for packet recovery, we develop an upper bound on the packet completion time of \ncrn.
% by taking into account the number of packets that each mobile device would like to receive for packet recovery. We develop an upper bound on the packet completion time of \ncrn.
% to exploit  develop an easy upper bound on the packet completion by taking into account the number of packets that each mobile device would like to receive for recovery.
\item We develop a network coding algorithm; \ncin, where packets are network coded in a way that they can be decoded immediately after they are received by their destination mobile devices. \ncin ~is crucial for multimedia applications with deadline requirements. Furthermore, we characterize the performance of \ncin, and we show through simulations that \ncin ~improves packet completion time significantly.
%which takes into account the types of packets that each mobile device would like to receive for packet recovery. We show that \ncin~ provides \textcolor{red}{instant decodability}.
% rather than just taking into account the number of packet each mobile device would like to receive.
%\vspace{-5pt}
\end{itemize}

The structure of the rest of this paper is as follows. Section \ref{sec:preliminaries} presents preliminaries and our problem statement. Sections \ref{sec:lower} and \ref{sec:upper} present lower and upper bounds on the performance of our network coding schemes, respectively. Section \ref{sec:simulation} presents simulation results. Section \ref{sec:related} presents the related work. Section \ref{sec:conclusion} concludes the paper.

\section{\label{sec:preliminaries}Preliminaries \& Problem Statement}
We consider a setup with $N$ cooperative mobile devices (nodes), where $\Nset$ is the set of devices in our system with $N=|\Nset|$. These devices are within close proximity of each other, so they are in the same transmission range. Note that the cooperative mobile devices in $\Nset$ are interested in receiving packets $p_m$ from set $\Mset$, \ie $p_m \in \Mset$ and $M = |\Mset|$.

Our system model consists of two stages. In the first stage, all packets are broadcast to all devices via cellular links. During the first stage, mobile devices may receive partial content due to packet losses over the cellular broadcast link.  Thus, after the first stage, the set of packets that mobile device $n \in \Nset$ has successfully received is $\Hset_n$, and is referred to as {\em Has} set of device $n$. The set of packets that are lost in the first stage at mobile device $n$ is referred to as {\em Wants} set of device $n$ and denoted by $\Wset_n$. In this paper, we assume that all mobile devices are interested in receiving all packets in $\Mset$. Thus, the following equality holds; $\Wset_n = \Mset \setminus \Hset_n$. Furthermore, we define the set $\Mset_c$ as $\Mset_c=\bigcap_{n \in \Nset} \Wset_n$. Note that the packets in $\Mset_c$ are not received by any devices in the local area during the first stage.

In the second stage, missing packets are recovered by utilizing both cellular and local area links. In particular, a mobile device may receive two recovery packets; one from cellular and another from local area link, simultaneously. Exploiting multiple interfaces has potential of improving throughput. Moreover, employing network coding further improves throughput in this setup. However, it is crucial to determine which network coded packets should be transmitted over cellular and local area links in this stage. This is an open problem and the focus of this paper. In particular, in this paper, we propose network coding algorithms for multiple interfaces (NCMI) to recover missing packets in the second stage.\footnote{\scriptsize We assume that network coded packets can be transmitted without any loss in the second stage. Note that our focus in this work is to understand the performance of network coding for cooperative mobile devices with multiple interfaces in a nutshell. Our approach in this paper is complementary to previous work \cite{salim_broadcast}, \cite{RouayhebITW10}, \cite{RouayhebISIT10}, and could be extended to include packet losses in the second stage.} Namely, we propose \ncrn ~for batch-based network coding and \ncin ~for instantaneously decodable network coding.

The integral part of our work is to analyze the throughput performance of \ncrn ~and \ncin. We consider the packet completion time as a performance metric, which is defined as follows:
\begin{definition}
\label{def:completion}
%\caption{Completion Time}
Packet completion time $T$ is the number of transmission slots in the second stage that are required for all mobile devices to decode all packets in their {\em Wants} sets.
\end{definition}

{\em Assumptions:} We assume, without loss of generality, that for each packet $p_m \in \Mset$, there is at least one mobile device that wants packet $p_m$. In other words, $\forall p_m \in \Mset, \exists n \in \Nset$ such that $p_m \in \Wset_n$. This assumption does not violate generality, because packets that are not wanted by any of the devices could be removed from $\Mset$.

\section{\label{sec:lower} Lower Bound on $T$}
In this section, we develop a lower bound on the packet completion time when any network coding algorithm is employed by cooperative mobile devices with multiple interfaces.
\begin{proposition} \label{th:lowerbound}
The packet completion time when network coding is employed by cooperative mobile devices with multiple interfaces is lower bounded by:
\begin{equation} \label{eq:lower}
T \geq \lceil \max(|\Mset_c|,\frac{1}{2}{\max_{n \in \Nset} |\Wset_n|}) \rceil.
\end{equation}
\end{proposition}
{\em Proof:} Each mobile device $n \in \Nset$ should receive at least $|\Wset_n|$ packets to be able to decode all the packets in its {\em Wants} set, $\Wset_n$. Therefore, the minimum number of packet transmissions is greater than or equal to $\max_{n \in \Nset} |\Wset_n|$. Since we have two packet transmissions at each transmission slot; one via cellular links and the other via local area links, the minimum completion time is $\frac{1}{2}{\max_{n \in \Nset} |\Wset_n|}$ in the best case scenario. On the other hand, since the packets in $\Mset_c$ can only be sent through the cellular link, the minimum completion time should be larger than $|\Mset_c|$. Thus, the completion time is bounded by $T \geq \max(|\Mset_c|,\frac{1}{2}{\max_{n \in \Nset} |\Wset_n|})$. Furthermore, since the completion time can only have an integer value, the completion time is lower bounded by $\lceil \max(|\Mset_c|,\frac{1}{2} {\max_{n \in \Nset} |\Wset_n|}) \rceil$. This completes the proof. \hfill $\blacksquare$

Note that the lower bound on $T$ in (\ref{eq:lower}) considers the best case scenario and characterizes the performance of network coding for this scenario. However, the actual completion time may be larger than the lower bound provided in (\ref{eq:lower}), so we develop \ncrn ~and \ncin ~in the next section, and characterize their upper bounds.

%we characterize the upper bounds on $T$ in the next section.

\section{\label{sec:upper} NCMI and Upper Bounds on $T$}
%In this section, we develop network coding algorithms for cooperative wireless networks with multiple interfaces (NCMI) and provide upper bounds on their packet completion time. In particular, we develop two network coding algorithms; \ncrn, which uses batch-based network coding, and \ncin, which provides instant decodability guarantee. %\textcolor{red}{We then compare the upper bounds obtained from \ncrn~ and \ncin.} % and analyze their completion times.

\subsection{\ncrn}
\subsubsection{Algorithm Description}
As we mentioned earlier in Section~\ref{sec:preliminaries}, our system model consists of two stages. In the first stage, all packets are broadcast to all devices via cellular links without network coding. In the second stage, both cellular and local area links are utilized simultaneously and network coding is employed. In particular, both the source and a local area node transmit network coded packets simultaneously at every transmission slot until there is no missing packet in the local area. Next, we explain how network coding is performed by the source and in the local area.

The source node (i) determines the missing packets in all mobile devices in the local area, (ii) transmits linear combinations of these packets (using random linear network coding over a sufficiently large field) over cellular links. These network coded packets are innovative and beneficial for any node $n$ for which $|\Hset_n| \le M$, because these network coded packets carry information about all missing packets in the local area. After each transmission, if the received packet is innovative for node $n$, it is inserted into $\Hset_n$ set. The procedure continues until each node $n$ receives $|\Wset_n|$ innovative packets.

On the other hand, in the local area, a mobile device $n_{\max}$ with the largest {\em Has} set; $n_{\max}=\arg \max_{n \in \Nset} |\Hset_n|$ is selected as the transmitter at each transmission slot. If there are multiple of such devices, one of them is selected randomly. The transmitter linearly combines all packets in its {\em Has} set, $\Hset_{n_{\max}}$, and broadcasts the network coded packet to all other mobile devices in the local area.
%This network coded packet is beneficial to any device $n'$ in the local area as long as $\Wset_{n'} \setminus \Mset_c$ is not an empty set.
After each transmission, if the received packet has innovative information for node $n$, the received packet is inserted into the {\em Has} set of node $n$. Note that the network coded packets that include packets from $\Mset_c=\bigcap_{n \in \Nset} \Wset_n$ can only be transmitted from the source, since these packets do not exist in the local area. Therefore, the local area devices stop transmitting network coded packets if each node $n$ receives (i) $|\Wset_n|-|\Mset_c|$ innovative packets from the local area, or (ii) $|\Wset_n|$ innovative packets from both the source and local area devices. Note that in \ncrn, there might exist more than one device that have the same set of network coded packets in their {\em Has} sets. In this case, without loss of generality, we consider these devices as one device to make network coding decisions.
% into a single user with the same {\em Has} and {\em Wants} sets of one of them. Also note that, at each transmission slot, there are at least two users with different informative packets in their {\em Has} sets unless each node $n$ receives $|\Wset_n|-|\Mset_c|$ informative packets from the local area or $|\Wset_n|$ informative packets from both local area and the source.}
%Note that the network coded packet is also not beneficial to the transmitting node $n$. Thus, in the second transmission slot of a time-frame, a device which has a packet beneficial for node $n$ is selected for transmission. This device transmits random linear combination of all packets in its {\em Has} set. Next, we explain our algorithm's operation via an example.}
%At each transmission slot, this procedure is repeated until there is no missing packets in the local area.

\begin{example}
Let us consider three mobile devices with the {\em Wants} sets; $\Wset_A=\{p_1,p_2,p_3\}, \Wset_B=\{p_1,p_4,p_5\}, \Wset_C=\{p_1,p_6,p_7\}$. Our algorithm combines $p_1, \ldots, p_7$ at the source, and transmits the network coded packet to the local area devices in the first slot. This transmitted packet is beneficial to all nodes in the local area as it carries information about all missing packets. Meanwhile, in the local area, a device with the largest {\em Has} set is selected. Since there is equality in this example ($|\Hset_A|=|\Hset_B|=|\Hset_C|=4$), one device is selected randomly, let us say device $A$. Device $A$ transmits linear combinations of $p_4, p_5, p_6$. Note that this network coded packet is beneficial to both device $B$ and $C$ as it carries information about their missing packets. Thus, in the first slot, the source transmits linear combination of $p_1, \ldots, p_7$ and device $A$ transmits the linear combination of $p_4, p_5, p_6$ simultaneously. In the next slot, the source transmits another linear combination of $p_1, \ldots, p_7$, while either $B$ or $C$ transmits a network coded packet as the sizes of their {\em Has} sets are $6$ (the sizes of their {\em Has} sets are increased by two because of the first transmissions) and the size of node $A$'s {\em Has} set is $5$ (the size of its {\em Has} set is increased by one, because it was the transmitter in the first transmission.). The same procedure is repeated at every slot until each node receives $3$ innovative packets. Next, we characterize how long it takes until all missing packets are recovered; \ie the completion time; $T$.
%
%In the first transmission slot, packet $p_1$ is sent from the base station and $p_5+p_7$ is sent from mobile device $A$. In the second transmission slot, $p_2+p_4+p_6$ is sent from the base station and $p_3$ is sent from mobile device $B$.
%
%By using linear network coding, all mobile devices can be satisfied in $\max(|\Mset_c|,\frac{2}{3}\max_{n \in \Nset} |\Wset_n|)=2$ transmission slots. In the first transmission slot, packet $p_1$ is sent from the base station and $p_5+p_7$ is sent from mobile device $A$. In the second transmission slot, $p_2+p_4+p_6$ is sent from the base station and $p_3$ is sent from mobile device $B$.
\hfill $\Box$
\end{example}

\subsubsection{Upper Bound on $T$}
\begin{theorem} \label{th:easy_upper_bound}
Packet completion time; $T$ when \ncrn ~is employed by cooperative mobile devices with multiple interfaces is upper bounded by
% is developed by taking into account the number of packets that each device would like to receive (\ie $\Mset_c$ and $\Wset_n$) as it follows:
\begin{equation} \label{eq:b2}
T \leq \lceil \max(|\Mset_c|,\frac{1}{3}(\max_{n \in \Nset} |\Wset_n|+\min_{n \in \Nset} |\Wset_n|), \frac{1}{2} \max_{n \in \Nset} |\Wset_n|) \rceil.
\end{equation}
\end{theorem}
{\em Proof:} The proof is provided in Appendix A.\hfill $\blacksquare$

\subsection{\ncin}
In this section, we develop \ncin, where packets are network coded in a way that they can be decoded immediately after they are received by their destination mobile devices. \ncin ~is crucial for multimedia applications with deadline requirements.
%which takes into account the types of packets (rather than just their numbers as in \ncrn), that each mobile device would like to receive for recovery. \textcolor{red}{We show that with this approach \ncin~ provides instant decodability}.
%We develop a network coding algorithm; \ncin, where packets are network coded in a way that they can be decoded immediately after they are received by their destination mobile devices. \ncin is crucial for multimedia applications with deadline requirements. Furthermore, we characterize the performance of \ncin, and we show through simulations that \ncin~ improves packet packet completion time significantly.

\subsubsection{Algorithm Description}
The key idea behind \ncin ~is to sort packets in $\bigcup_{n \in \Nset} \Wset_n$ based on their differences from the view point of the cellular and the local area links, and create the following sets; $\Mset_c$, $\Mset_l$, and $\Mset_d$. These sets consist of (possibly) network coded packets. Our grouping algorithm is provided in Algorithm \ref{al:Groups}. %, and we provide the details about Algorithm \ref{al:Groups} in Appendix C.

\begin{algorithm}[h]
\caption{Grouping the packets in the {\em Wants} Sets}
\label{al:Groups}
\begin{algorithmic}[1]
%\STATE Reorder packets in $\Mset$ so that its first $|\Mset_c|$ packets are from $\Mset_c$.
\begin{scriptsize}
\FOR{any packet $p_m$ in $\Mset$}
\STATE Define vector $\boldsymbol v_m$ with size $\Nset$. Each element of the vector $\boldsymbol v_m$ is set to NULL initially; \ie $\boldsymbol v_m[n] = NULL$, $\forall n \in \Nset$.
\FOR{any device $n$ in $\Nset$}
\IF {$p_m$ is wanted by node $n$}
\STATE $\boldsymbol v_m[n] = p_m$ and $p_m$ is removed from the {\em Wants} set $\Wset_n$.
\ENDIF
\ENDFOR
\IF {there exists a vector $\boldsymbol v_{m'}$, $m' < m$ satisfying either (i) $\boldsymbol v_{m'}[n] = NULL$ AND $\boldsymbol v_{m'}[n] \neq NULL$, OR (ii) $\boldsymbol v_{m'}[n] \neq NULL$ AND $\boldsymbol v_{m'}[n] = NULL$, for any $n$ }
\STATE Replace $\boldsymbol v_{m'}$ with $\boldsymbol v_{m'} + \boldsymbol v_m$ and delete $\boldsymbol v_{m}$. (Note that $p_m + NULL = p_m$ for any $m$)
%\STATE Select the packet with the smallest sequence number from the {\em Wants} set $\Wset_n$. Let us say this packet is $p_l$. Then,   $\boldsymbol v_k[n] = p_l$ and $p_l$ is removed from the {\em Wants} set $\Wset_n$
\ENDIF
\ENDFOR
\STATE Each element of $\Mset_c$ is constructed by network coding all packets in a vector if the vector satisfies the following condition: the elements of the vector should be the same.
\STATE Determine the node with minimum {\em Wants} set as $n^*$ where $n^*=\arg \min_{n \in \Nset} |\Wset_n|$. Each element of $\Mset_d$ is constructed by network coding all packets in a vector if the vector satisfies the following condition: the  vector's $n^*$th element should be NULL.
\STATE Construct $\Mset_l$ using the remaining vectors.
\end{scriptsize}
\end{algorithmic}
\end{algorithm}
%\vspace{-5pt}

The packets in $\Mset_c$ can only be transmitted from the base station (source node), because they do not exist in the local area, so packet recovery in the local area is not possible. Thus, the source node transmits packets from $\Mset_c$ without network coding.

The packets in $\Mset_d$ are network coded packets that can be transmitted by a single transmission from both the source or a local area device. Note that network coded packets can be transmitted by a single transmission from the cooperating devices by selecting $n^*$ (the device with the minimum size of {\em Wants} set) as the transmitter. Thus, $\Mset_d$ is constructed by taking into account which instantly decodable network coded packets can be generated from the device with the minimum size of {\em Wants} set.

The packets in $\Mset_l$ are the rest of the network coded packets that can be created and decodable by all mobile devices. We note that it takes one transmission from the base station and maximum of two transmissions from the cooperating devices to send each packet in $\Mset_l$. The reason for this is that the base station has all of the packets, so that any packet combination is available and can be broadcast to all devices. On the other hand, in the local area, there is no guarantee that the network coded packet in $\Mset_l$ can be created and transmitted. Thus, if there does not exist a device that can generate the network coded packet, then a part of the network coded packet is created and transmitted. In this case, two transmissions are necessary and sufficient to transmit the content of the network coded packets. We explain this fact as well as the properties of $\Mset_c, \Mset_l,$ and $\Mset_d$ via an example. The properties of the sets $\Mset_c, \Mset_l,$ and $\Mset_d$ are also provided in Table~\ref{table:diff}.

%one mobile device is selected as the transmitter, and it transmits the network coded packet, which is beneficial for all other devices except itself and the users that want the same packet as the sender. Thus, another transmission is necessary to satisfy the transmitter device's packet requirement. Therefore, it may take up to two transmissions for the cooperating devices to send each packet in $\Mset_l$.

% Next, we explain these sets via an example.

\begin{table}  % is used to refer this table in the text
\caption{The required number of transmissions to transmit each packet in $\Mset_c, \Mset_l,$ and $\Mset_d$ via cellular and local area links.} % title of Table
%\vspace{-10pt}
\centering % used for centering table
\begin{tabular}{c c c } % centered columns (4 columns)
\hline
%\multicolumn{1}{c}{} & \multicolumn{3}{|c|}{Total Distortion}\\
%\hline\hline %inserts double horizontal lines
\scriptsize Set & \scriptsize Cellular Link & \scriptsize Local Area Link \\ [0.5ex]
\hline % inserts single horizontal line
\scriptsize $\Mset_c$ & \scriptsize $1$ & \scriptsize N/A \\ % inserting body of the table
\scriptsize $\Mset_l$ & \scriptsize $1$ & \scriptsize $\leq 2$ \\
\scriptsize $\Mset_d$ & \scriptsize $1$ & \scriptsize $1$ \\
%News & $19.5\%$ & $20.7\%$ & $28.8\%$ & $3.8\%$ & $4.6\%$ & $4.9\%$ % [1ex] adds vertical space
\end{tabular} \label{table:diff}
\vspace{-10pt}
\end{table}

\begin{example}\label{ex:group}
Let us assume that there are three mobile devices with the {\em Wants} sets; $\Wset_A=\{p_1,p_2,p_3,p_5\}$, $\Wset_B=\{p_1,p_2,p_3,p_6,p_8\}$, $\Wset_C=\{p_1,p_2,p_4,p_7,p_9,p_{10}\}$. Note that $\Mset= \bigcup_{n \in \Nset} \Wset_n$ and $\Hset_n = \Mset \setminus \Wset_n$ for $n \in \{A,B,C\}$.
According to Algorithm~\ref{al:Groups}, $\Mset_c=\{p_1,p_2\}$; $p_1$ and $p_2$ can only be sent from the base station, because they are not available in any {\em Has} sets of the mobile devices. $\Mset_l$ is equal to $\Mset_l=\{p_3 + p_4, p_5 + p_6 + p_7\}$. Each packet in $\Mset_l$ can be sent either by a single transmission from the base station or by one or two transmissions from the cooperating devices. Let us consider the transmission of $p_3 + p_4$. The base station can broadcast this packet directly. On the other hand, in the local area, two transmissions are required: (i) device $A$ sends $p_4$, which is decodable by $C$, and (ii) device $C$ sends packet $p_3$, which is decodable by $A$ and $B$. As can be seen, transmission of packets in $\Mset_l$ may take one or two transmission slots if they are transmitted from cooperating devices. The set $\Mset_d$ is equal to $\Mset_d=\{p_8 + p_9,p_{10}\}$. The packets in $\Mset_d$ can be sent from either the base station or from device $A$ in the local area, and for both cases one transmission is sufficient. %Packet $p_8 + p_9$ is decodable for users $2$ and $3$ and packet $p_{10}$ is decodable for user $3$.
\hfill $\Box$
\end{example}

After packets are grouped into sets $\Mset_c$, $\Mset_l$, and $\Mset_d$, each element in $\Mset_c \cup \Mset_l \cup \Mset_d$ becomes a (network coded) packet that is instantly decodable for all or a subset of mobile devices.
%When we transmit all the packets in $\Mset_c \cup \Mset_l \cup \Mset_d$, mobile devices receive all the packets in their {\em Wants} sets, hence transmissions and packet recovery are completed. We give an example of how $NCmI$ constructs sets  $\Mset_c$, $\Mset_l$, and $\Mset_d$, and network coded packets in the following example.
In particular, after packets are grouped, \ncin ~transmits two packets simultaneously at each transmission slot; one from the base station and another from one of the cooperating devices. The base station starts sending the packets from $\Mset_c$. After all packets in $\Mset_c$ are transmitted from the base station, the remaining packets in $\Mset_l$ are transmitted, and finally the remaining packets in $\Mset_d$ are transmitted. Meanwhile, the cooperating mobile devices start sending the packets in $\Mset_d$. After all the packets in $\Mset_d$ are transmitted, the remaining packets in $\Mset_l$ are transmitted. Next, we describe how \ncin ~determines mobile devices to transmit packets from $\Mset_l$ and $\Mset_d$.

In order to transmit a network coded packet from $\Mset_l$ using local area links, \eg $p_5 + p_6 + p_7$ in Example~\ref{ex:group}, \ncin ~first looks for a device that can transmit the network coded packet. If there exists such a device, the network coded packet is transmitted. However, if there is no such device in the local area, as in the case for $p_5 + p_6 + p_7$ in Example~\ref{ex:group}, then a device is selected randomly. This device can create a partial network code; \eg if device $A$ is selected in Example~\ref{ex:group}, it transmits the partial network coded packet $p_6 + p_7$ to $B$ and $C$. Then, another device is selected to send the transmitting device's packet requirement; \eg device $B$ transmits $p_5$ to device $A$. Note that we construct $\Mset_l$ such that a network coded packet is transmitted in the local area in one or two transmissions. On the other hand, packets in $\Mset_d$ can be transmitted in the local area from node with the smallest {\em Wants} sets by a single transmission, because according to Algorithm \ref{al:Groups}, it is guaranteed that these packets are available in the {\em Has} set of this device. In the next section, by taking into account the relative sizes of the sets $\Mset_c, \Mset_l$ and $\Mset_d$, we will develop an upper bound on the completion time of \ncin.

\subsubsection{Upper Bound on $T$}
%Now, we provide an upper bound on the completion time of $NCmI$.

\begin{theorem}\label{theorem:better_upper}
An upper bound on the packet completion time when \ncin ~is employed by cooperative mobile devices with multiple interfaces is:
\begin{align} \label{eq:better_upper}
& T \leq  \lceil \mathlarger{\max}(|\Mset_c|,(\frac{1}{3}(2 \min_{n \in \Nset} |\Wset_n|+|\Mset_d|), \frac{1}{2}(\min_{n \in \Nset} |\Wset_n| \nonumber \\
& +|\Mset_d|)) \rceil.
\end{align}
\end{theorem}
{\em Proof:} The proof is provided in Appendix B. \hfill $\blacksquare$
\vspace{-5pt}
\subsection{\ncrn ~versus \ncin}
\vspace{-5pt}
In this paper, we proposed two network coding algorithms for cooperative mobile devices with multiple interfaces; \ncrn ~as a batch-based network coding, and \ncin ~as an instantly decodable network coding. These algorithms bring different strengths for different applications. For data intensive applications, \ncrn ~is more applicable as it improves throughput significantly and more as compared to \ncin. For multimedia applications with deadline constraints \cite{CA_IDNC}, \ncin ~is more applicable as it provides instant decodability. The next proposition shows that \ncrn ~further improves throughput as compared to \ncin.
\begin{proposition} \label{th:difference}
The upper bound of \ncrn ~provided in (\ref{th:easy_upper_bound}) is tighter as compared to the upper bound of \ncin ~provided in (\ref{eq:better_upper}).
\end{proposition}
{\em Proof:} By using the fact that $\min_{n \in \Nset} |\Wset_n|+|\Mset_d|=|\Mset_c|+|\Mset_l|+|\Mset_d| \ge \max_{n \in \Nset} |\Wset_n|$, we have the following inequalities:

\begin{align}
&\frac{1}{3}(2 \min_{n \in \Nset} |\Wset_n|+|\Mset_d|) \ge \frac{1}{3}(\min_{n \in \Nset} |\Wset_n|+\max_{n \in \Nset} |\Wset_n|)\\
&\frac{1}{2}(\min_{n \in \Nset} |\Wset_n| +|\Mset_d|) \ge \frac{1}{2}(\max_{n \in \Nset} |\Wset_n|)
\end{align}

By using the above inequalities, it is easy to show that the upper bound obtained from \ncrn ~is larger than the upper bound obtained from \ncin. \hfill $\blacksquare$

Even though \ncrn ~outperforms \ncin, the performances of \ncin ~and \ncrn ~are close to each other and also close to the lower bound provided in (\ref{eq:lower}) as we show via simulations in the next section. %Furthermore, \ncin~ provides instant decodability, which is crucial for multimedia applications with deadline constraints \cite{CA_IDNC}.
%The result mentioned in proposition~\ref{th:difference} is expected, because random linear network coding has the best performance among all network coding schemes including instantly decodable network coding. However, IDNC is more efficient in applications such as multimedia, where the content should be played out before a hard deadline.
%\input{gap}

\section{\label{sec:simulation}Simulation Results}
We implemented our proposed schemes: \ncrn ~and \ncin, and compared their completion time performance with: (i) the {\em Lower Bound}, in (\ref{eq:lower}), (ii) their {\em Upper Bounds} provided in (\ref{eq:b2}) and (\ref{eq:better_upper}),
(iii) {\em No-NC}, which is a no network coding scheme, but using cooperation and multiple interfaces,
(iv) {\em Single-Interface NC, via Cellular Links}, which uses a single interface, namely cellular links, and uses batch-based network coding,
(v) {\em Single-Interface NC, via Local Area Links}, which uses mainly local area links, and uses batch-based network coding. Note that packets in $\Mset_c$ are requested from the source node via the cellular links in {\em Single-Interface NC, via Local Area Links} scheme.
%which the missing packets in $\Mset_c$ are recovered by sending these packets from the source and the remaining missing packets are recovered by sending network coded packets through the local area links.
%This bound is achieved when the missing packets in $\bigcup_{n \in \mathcal{N}} W_n$ are sent from the cellular and local area links without network coding.
We consider a topology shown in Fig.~\ref{fig:intro_example}(b) with $N=5$ mobile devices and for different number of packets and loss probabilities. In our simulation results, bounds are plotted using dashed curves, while the real simulation results are plotted using the solid curves.

\begin{figure}[t!]
\centering
\vspace{-5pt}
\subfigure[\scriptsize Completion time vs. number of packets]{ \scalebox{.5}{\includegraphics{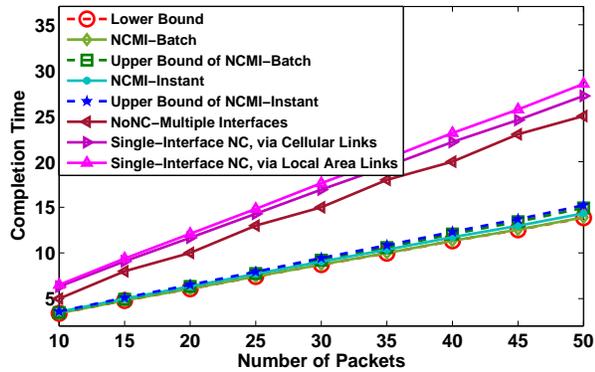}} }
\subfigure[\scriptsize  Completion time vs. loss probability]{ \scalebox{.5}{\includegraphics{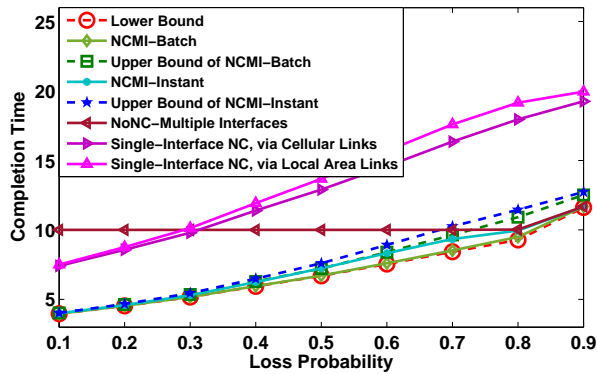}} }
\vspace{-8pt}
\caption{\scriptsize The completion time performance of \ncin ~and \ncrn ~as compared to their lower and upper bounds as well as baselines.
%, {\em Lower Bound}, {\em Easy Upper Bound}, {\em Better Upper Bound}, and {\em Trivial Bound}.
}
\vspace{-15pt}
\label{fig:sumulations}
\end{figure}

{\bf Completion time vs. number of packets:} Fig. \ref{fig:sumulations}(a) shows the completion time for different number of packets. In this setup, each device selects its loss probability uniformly from $[0.3, 0.5]$, and looses packets according to the selected loss probability. Note that the number of lost packets is equal to $M=|\bigcup_{n \in \mathcal{N}}W_n|$ in Fig. \ref{fig:sumulations}(a). As seen, \ncin ~and \ncrn ~improve the completion time significantly as compared to the single-interface systems and No-NC. This shows the effectiveness of using multiple interfaces compared to the single-interface systems. \ncrn ~and \ncin ~are slightly better than their upper bounds for larger number of lost packets, because the upper bounds give the worst case performance guarantee for \ncrn ~and \ncin, respectively. Finally, the completion times of \ncrn ~and \ncin ~and their upper bounds are very close to the lower bound, which demonstrates the effectiveness of our network coding design for cooperative devices with multiple interfaces. As expected, \ncrn ~outperforms \ncin, but \ncin ~also significantly improves packet completion time.

{\bf Completion time vs. loss probability:} Fig. \ref{fig:sumulations}(b) presents the completion time for different loss probabilities when $M=20$. In this setup, the loss probability is the same for all mobile devices. As seen, \ncrn ~and \ncin ~significantly improve the completion time as compared to single-interface systems and No-NC scheme, and shows very close performance as compared to the lower bound. %In The x-axis shows the loss probability which is the same for all users $n \in \mathcal{N}$. As shown in the figure, the completion time increases with an increase in the loss probability.

{\bf Computational Complexity:} The complexity of \ncrn ~is linear with the number of nodes in the local area and the number of packets. In particular, a node with the largest {\em Has} set is selected for transmission at each slot with complexity $O(N)$. Then, packets are network coded with complexity $O(M)$. The complexity of \ncin ~is polynomial time with $O(MN+M^2)$. In particular, Algorithm \ref{al:Groups} constructs vectors by checking all packets and devices in the local area with complexity $O(MN)$. The constructed vectors will be merged with the complexity of $O(M^2)$. This computational complexity, by also taking additional steps such as dividing a file into smaller sets of $M$ packets, makes both \ncrn ~and \ncin ~applicable for practical deployment.

\section{\label{sec:related} Related Work}
{\em Network Coding for Single-Interface Systems:} The performance of network coding has been evaluated for single-interface systems in literature. %Capacity region characterization of single-hop broadcast topology with network coding and erasure channels with feedback is considered in \cite{cc_wang}.
The problem of minimizing the number of broadcast transmissions required to satisfy all nodes is considered in \cite{salim_broadcast}, and the bounds for completion time are developed. A deterministic linear network coding algorithm that minimizes the number of broadcast transmissions is considered in  \cite{parastoo_broadcast}.  Minimization of the completion delay while broadcasting instantly decodable network coding packets has been considered in \cite{sorourICC}. The problem of recovering the missing content using cooperative data exchange utilizing local area connections is considered in \cite{RouayhebITW10} and \cite{RouayhebISIT10}, and the lower and upper bounds on the minimum number of transmissions are developed. Deterministic algorithms for the cooperative data exchange problem with polynomial time are designed in \cite{RouayhebISIT11} and \cite{SprintsonQShine10}.  As compared to this line of work, we consider cooperative mobile devices with multiple interfaces, and develop a network coding scheme for this setup.

{\em Network Coding for Multiple-Interface Systems:} Network coding has been employed in the previous work for devices with multiple interfaces. Wireless video broadcasting with P2P error recovery is proposed by Li and Chan \cite{micro18}. An efficient scheduling approach with network coding for wireless local repair is introduced by Saleh et al. \cite{micro19}. Another body of work \cite{micro20}, \cite{micro21}, \cite{micro22} proposes systems where there are a base station broadcasting packets and a group of smartphone users helping each other to correct errors. Compared to prior work \cite{micro18}, \cite{micro19}, \cite{micro20}, \cite{micro21}, \cite{micro22}, where each phone downloads all the data, and the local links are used for error recovery, our scheme jointly utilizes multiple interfaces and analyzes the performance of network coding in such a setup.

Simultaneous operation of multiple interfaces and employing network coding for this setup has also been considered in the previous work; \cite{microcast}, \cite{microcast_allerton}, \cite{AliParan11}, where multiple interfaces are used to improve the download rate at each mobile device. As compared to this line of work, we consider how efficient network coding algorithms can be developed with provable performance guarantees for cooperative mobile devices with multiple interfaces, instead of using existing network coding algorithms.

\section{\label{sec:conclusion}Conclusion}
%In this paper, we investigated the performance of network coding for cooperative mobile devices with multiple interfaces, and developed a network coding scheme for this setup. In particular,
In this paper, we considered a scenario where a group of mobile devices is interested in the same content, but each device has a partial content due to packet losses over links. In this setup, mobile devices cooperate and exploit their multiple interfaces to recover the missing content. We developed network coding schemes; \ncrn ~and \ncin ~for this setup, and analyzed their completion time. Simulation results confirm that \ncrn ~and \ncin ~significantly reduce the completion time.

\section*{\label{appendix_upper_batch}Appendix A: Proof of Theorem~\ref{th:easy_upper_bound}}
In \ncrn, at each transmission slot, the transmitted packet from the cellular links is beneficial to all users and the transmitted packet from the local area links is beneficial to all except for the transmitter. Therefore, the size of {\em Has} set for each user is increased by two except for the user with the maximum size of {\em Has} set (the transmitter). Consider the two users $n_{min}$ and $n_{max}$ with the minimum and maximum size of {\em Has} set among all nodes at the beginning of \ncrn ~algorithm; $n_{\max}=\arg \max_{n \in \Nset} |\Hset_n|$ and $n_{\min}=\arg \min_{n \in \Nset} |\Hset_n|$. The upper bound on $T$ is equal to the maximum number of transmission slots required to satisfy node $n_{min}$. Therefore, we next analyze the required number of transmission slots to satisfy node $n_{min}$.
In the first transmission slot, $n_{max}$ is selected as the transmitter so the size of $\Hset_{n_{min}}$ is increased by two and the the size of $\Hset_{n_{max}}$ is increased by one. It takes at most $|\Hset_{n_{max}}|-|\Hset_{n_{min}}|$ transmission slots that the size of $\Hset_{n_{max}}$ becomes equal to the size of $\Hset_{n_{min}}$. On the other hand, it takes at most $M-|\Hset_{n_{max}}|$ transmission slots that the size of $\Hset_{n_{max}}$ becomes equal to $M$. We consider two cases:

\begin{enumerate}

\item $(M- |\Hset_{n_{max}}|) \leq |\Hset_{n_{max}}|-|\Hset_{n_{min}}|$:

After at most $M-|\Hset_{n_{max}}|$ transmission slots, the size of $\Hset_{n_{max}}$ becomes equal to $M$ and the size of $\Hset_{n_{min}}$ is still less than the size of $\Hset_{n_{max}}$. Therefore, in the next transmission slots $n_{max}$ is selected as the transmitter. Thus, the total number of transmission slots required to satisfy node $n_{min}$ is equal to $\frac{1}{2} \max_{n \in \Nset}|\Wset_n|$.

\item $(M- |\Hset_{n_{max}}|) \ge |\Hset_{n_{max}}|-|\Hset_{n_{min}}|$:

After at most $|\Hset_{n_{max}}|-|\Hset_{n_{min}}|$ transmission slots, the size of $\Hset_{n_{max}}$ becomes equal to the size of $\Hset_{n_{min}}$. Therefore, in the next transmission slots, $n_{min}$ and $n_{max}$ are selected as the transmitter alternatively and thus in every two transmission slots, the size of $\Hset_{n_{min}}$ is increased by three. This results in the total number of transmission slots to satisfy $n_{min}$ to be equal to $\frac{1}{3} (|\Wset_{n_{min}}|+|\Wset_{n_{max}}|)$ under the condition of this case.

\end{enumerate}

In addition, the number of transmissions cannot be less than $|\Mset_c|$. By considering this fact and the results from cases (i) and (ii), the upper bound in Theorem~\ref{th:easy_upper_bound} is obtained. This concludes the proof.

\section*{\label{appendix_upper_instant}Appendix B: Proof of Theorem~\ref{theorem:better_upper}}
We consider three conditions based on the relative sizes of the sets $\Mset_c$, $\Mset_l$ and $\Mset_d$ and then calculate the maximum completion time obtained from each of the conditions.

\begin{enumerate}
\item{$|\Mset_c| \geq (|\Mset_d|+2|\Mset_l|)$}\\
Under this condition, the base station starts sending the packets in $\Mset_c$; meanwhile, the cooperating nodes send the network coded packets in $\Mset_d$ and $\Mset_l$ respectively. After maximum of $|\Mset_d|+2|\Mset_l|$ transmission slots all the packets in $\Mset_d$ and $\Mset_l$ are sent by the cooperating nodes and $|\Mset_c|-(|\Mset_d|+2|\Mset_l|)$ packets are left from $\Mset_c$; it takes $|\Mset_c|-(|\Mset_d|+2|\Mset_l|)$ transmission slots for the base station to send these remaining packets. By summing the required number of transmission slots, the completion time under condition (1) is equal to:

\begin{equation}
T_{(1)}=|\Mset_c|
\end{equation}

\begin{example}
Let us consider three users with the {\em Wants} sets of $\Wset_A=\{p_1,p_2,p_3,p_4,p_5\}, \Wset_B=\{p_1,p_2,p_3,p_4,p_6,p_7\}, \Wset_C=\{p_1,p_2,p_3,p_4,p_6,p_8\}$. By using Algorithm \ref{al:Groups}, $\Mset_c=\{p_1,p_2,p_3,p_4\}, \Mset_l=\{p_5+2p_6\}, \Mset_d=\{p_7+p_8\}$. For this example, condition (1) is met; $|\Mset_c|=4>3=(|\Mset_d|+2|\Mset_l|)$. Accordingly, $4$ transmission slots are required; in the first transmission slot, $p_1$ is sent from the base station and at the same time $p_7+p_8$ is sent from user $1$. In the second transmission slot, $p_2$ is sent from the base station and $2p_6$ is sent from user $1$. In the third transmission slot, $p_3$ is sent from the base station and $p_5$ is sent from user $2$ (or user $3$). In the forth transmission slot, $p_4$ is sent from the the base station.
\hfill $\Box$
\end{example}

\item{$(|\Mset_d|+2|\Mset_l|) \geq \Mset_c \geq (|\Mset_d|-|\Mset_l|)$}\\
For this condition, we consider two cases of (i) $|\Mset_c| \leq |\Mset_d|$ and (ii) $|\Mset_c| \geq |\Mset_d|$.

In case (i), the base station starts sending the packets in $\Mset_c$; meanwhile $n^*$ (as the transmitter among the cooperating nodes) starts sending the packets in $\Mset_d$. Since $|\Mset_c| \leq |\Mset_d|$, after $|\Mset_c|$ transmission slots, all the packets in $\Mset_c$ have been transmitted by the base station and $|\Mset_d|-|\Mset_c|$ packets are left from $\Mset_d$. According to condition (2), $|\Mset_c|$ is greater than $(|\Mset_d|-|\Mset_l|)$. Therefore, in the next $|\Mset_d|-|\Mset_c|$ transmission slots, $n^*$ (as the transmitter among the cooperative nodes) sends the remaining packets in $\Mset_d$ and the base station transmits the network coded packets from $\Mset_l$. At last, $|\Mset_l|-(|\Mset_d|-|\Mset_c|)$ packets are left from $\Mset_l$; it takes maximum of $2/3(|\Mset_l|-(|\Mset_d|-|\Mset_c|))$ transmission slots by using both cooperating nodes and the base station to send these remaining packets. By summing the required number of transmission slots, the maximum completion time for case (i) is equal to $\frac{1}{3}(2|\Mset_l|+2|\Mset_c|+|\Mset_d|)$.

In case (ii), the base station starts sending the packets in $\Mset_c$; meanwhile $n^*$ (as the transmitter among the cooperating nodes) sends the packets in $\Mset_d$. After $|\Mset_d|$ transmission slots, all the packets in $\Mset_d$ have been transmitted by $n^*$ and $|\Mset_c|-|\Mset_d|$ packets are left from $\Mset_c$. In the next $|\Mset_c|-|\Mset_d|$ transmission slots, the base station sends the remaining packets in $\Mset_c$ and the cooperating nodes send $\frac{|\Mset_c|-|\Mset_d|}{2}$ packets from $\Mset_l$. At last, $|\Mset_l|-\frac{|\Mset_c|-|\Mset_d|}{2}$ packets are left from $|\Mset_l|$; it takes maximum of $\frac{2}{3}(|\Mset_l|-\frac{|\Mset_c|-|\Mset_d|}{2})$ transmission slots by using both cooperating nodes and the base station to send these remaining packets. By summing the required number of transmission slots, the completion time for case (ii) is equal to $\frac{1}{3}(2|\Mset_l|+2|\Mset_c|+|\Mset_d|)$.

Therefore, the maximum completion time under condition (2) is equal to::
\begin{equation}
\begin{split}
T_{(2)}&=\frac{1}{3}(2|\Mset_l|+2|\Mset_c|+|\Mset_d|)\\
&=\frac{1}{3}(2\min_{n \in \Nset} |\Wset_n|+|\Mset_d|).
\end{split}
\end{equation}

\begin{example}
Let us consider three users with the {\em Wants} sets of $\Wset_A=\{p_1,p_2,p_5,p_8\}, \Wset_B=\{p_1,p_3,p_6,p_9,p_{11}\}, \Wset_C=\{p_1,p_4,p_7,p_{10},p_{11}\}$. By using Algorithm \ref{al:Groups}, $\Mset_c=\{p_1\}, \Mset_l=\{p_2+p_3+p_4,p_5+p_6+p_7,p_8+p_9+p_{10}\}, \Mset_d=\{2p_{11}\}$. For this example, condition (2) is met; $(|\Mset_d|-|\Mset_l|)<|\Mset_c|<(|\Mset_d|+2|\Mset_l|)$. Accordingly, $3$ transmission slots are required; in the first transmission slot, $p_1$ is sent from the base station and at the same time $p_{11}$ is sent from user $1$. In the second transmission slot, $p_2+p_3+p_4$ is sent from the base station and $p_9+p_{10}$ is sent from user $1$. In the third transmission slot, $p_5+p_6+p_7$ is sent from the base station and $p_8$ is sent from user $2$.
\hfill $\Box$
\end{example}

\item{$|\Mset_c| \leq (|\Mset_d|-|\Mset_l|)$}\\
Under this condition, the base station starts sending the packets in $\Mset_c$; meanwhile $n^*$ (as the transmitter among the cooperating nodes) starts sending the packets in $\Mset_d$. After $|\Mset_c|$ transmission slots, all the packets in $\Mset_c$ have been sent by the base station and $|\Mset_d|-|\Mset_c|$ packets are left from $\Mset_d$. In the next $|\Mset_l|$ transmission slots, the base station sends all the packets in $\Mset_l$ and $n^*$ sends $|\Mset_l|$ packets from $\Mset_d$. At last, $|\Mset_d|-|\Mset_c|-|\Mset_l|$ packets are left from $\Mset_d$; it takes $\frac{|\Mset_d|-|\Mset_c|-|\Mset_l|}{2}$ transmission slots by using both cooperating nodes and the base station to send these remaining packets. By summing the required number of transmission slots, the completion time under condition (3) is equal to:

\begin{equation}
\begin{split}
T_{(3)}&=\frac{|\Mset_d|+|\Mset_c|+|\Mset_l|}{2}\\
&=\frac{|\Mset_d|+\min_{n \in \Nset} |\Wset_n|}{2}
\end{split}
\end{equation}

\begin{example}
Let us consider three users with the {\em Wants} sets of $\Wset_A=\{p_1,p_2\}, \Wset_B=\{p_1,p_3,p_5,p_6,p_9\}, \Wset_C=\{p_1,p_4,p_5,p_7,p_8,p_{10}\}$. By using Algorithm \ref{al:Groups}, $\Mset_c=\{p_1\}, \Mset_l=\{p_2+p_3+p_4\}, \Mset_d=\{p_5,p_6+p_7,p_8,p_9+p_{10}\}$. For this example, condition (3) is met; $|\Mset_c|<(|\Mset_d|-|\Mset_l|)$. Accordingly, $3$ transmission slots are required; in the first transmission slot, $p_1$ is sent from the base station and at the same time $p_9+p_{10}$ is sent from user $1$. In the second transmission slot, $p_2+p_3+p_4$ is sent from the base station and $p_8$ is sent from user $1$. In the third transmission slot, $p_5$ is sent from the base station and $p_6+p_7$ is sent from user $1$.
\hfill $\Box$
\end{example}

\end{enumerate}

By combining the completion time obtained from conditions (1), (2), and (3), the upper bound in Theorem \ref{theorem:better_upper} is achieved. This concludes the proof.

\end{document}